\documentclass[11pt]{article}
\usepackage{Blois,epsfig}

\def\be{\begin{equation}}
\def\ee{\end{equation}}
\def\bea{\begin{eqnarray}}
\def\eea{\end{eqnarray}}

\begin{document}
\vspace*{2cm}
\begin{center}
\Large{\textbf{XIth International Conference on\\ Elastic and Diffractive Scattering\\ Ch\^{a}teau de Blois, France, May 15 - 20, 2005}}
\end{center}

\vspace*{2cm}
\title{THE CURRENT STATUS AND PROSPECT OF THE TA EXPERIMENT}

\author{K. KASAHARA for the TA Collaboration%
\footnote{%
Author list:
H.Kawai, T.Nunomura,  S.Yoshida(Chiba University), H.Yoshii(Ehime University),
K.Tanaka(Hiroshima City University), 
F.Cohen, M.Fukushima, N.Hayashida, M.Ohnishi, H.Ohoka, S.Ozawa, N.Sakurai, H.Sagawa, T.Shibata, H.Shimodaira, M.Takeda, A.Taketa, M.Takita, H.Tokuno, R.Torii, S.Udo(ICRR, University of Tokyo), H.Fujii,T.Matsuda, M.Tanaka, H.Yamaoka(Institute of Particle and Nuclear Studies, KEK),
K.Hibino(Kanagawa University), T.Benno, M.Chikawa(Kinki University),
T.Nakamura(Kochi University), M.Teshima(Max-Planck-Institute for Physics), 
K.Kadota(Musashi Institute of Technology), 
Y.Uchihori(National Institute of Radiological Sciences),
Y.Hayashi, S.Kawakami, K.Matsumoto,Y.Matsumoto, T.Matsuyama, S.Ogio, 
A.Ohshima, T.Okuda(Osaka City University),
D.R.Bergman, G.Hughes, S.Stratton, G.B.Thomson(Rutgers University), 
N.Inoue, Y.Wada(Saitama University), 
M.Fukuda, T.Iguchi, F.Kakimoto, R.Minagawa, Y.Tameda, Y.Tsunesada( Tokyo Institute of Technology),
J.W.Belzq(University of Montana \& University of Utah),
J.A.J.Matthews(University of New Mexico),
T.Abu-Zayyad, R.Cadys, Z.Cao, P.Huentemeyer, C.C.H.Jui, K.Martens, J.N.Matthews, J.D.Smith, P.Sokolsky, R.W.Springer, S.B.Thomas, L.R.Wiencke(University of Utah),
T.Doyle, M.J.Taylor,  V.B.Wickwar, T.D.Wilkerson(Utah State University),
K.Hashimoto, K.Honda, T.Ishii, T.Kanbe(Yamanashi University)
}}
\address{Department of Electronic and Information Systems, Faculty of Systems Engineering, 307 
Fukasaku, Minuma-ku, Saitama-shi,  Saitama, Japan}

\maketitle\abstracts{
The Telescope Array (TA) experiment is designed 
to observe  cosmic-ray-induced  air showers at extremely high energies.
It is being  deployed in a desert of Utah, USA;  an array of 3 m$^2$ scintillation counters
will be distributed over  760 km$^2$  and 3 sets of air fluorescence telescopes will be
placed in the perimeter of the array. It's primary purpose is to make a decisive measurement of the cosmic ray spectrum in the GZK cutoff region.
We expect the first data from the TA in the spring of 2007.  As  its unique features are included
1) hybrid measurement  planned down to 10$^{17.5}$ eV,  2) calibration
of fluorescence detection by using artificial air showers generated by an electron linac,
3) interaction model calibration by the LHC.
}

\section{Introduction}
\subsection{Why Extremely High Energy Cosmic Rays (EHECRs)?}

EHE protons are expected to reach the Earth after  
traveling  distant extragalactic space  and therefore lose 
energy by interacting with the cosmic microwave background (CMB).  Lorentz invariance tells us
that the interaction takes place over $\sim 10^{20}$ eV for protons;  this will result in significant energy loss
of  protons and show up in the energy spectrum of EHE protons as a rather sharp cutoff over
$10^{20}$eV.   This  was first predicted by Greisen, Zatsepin and Kuzmin\,\cite{gzk}
and called the GZK cutoff.
Should there be  cosmic rays well above $10^{20}$ eV (= super-GZK events),
it implies their origin be most probably  within  50 Mpc from the Earth.

What  interesting is that the energy spectrum measured by AGASA
 shows no indication of the GZK cutoff.\cite{agasa}
They recorded 11 super-GZK cosmic rays in 13 years of operation. 
Moreover, in the sample of 59 events above $10^{19.6}$ eV, 
13 events were found to form a cluster of 2 or 3 events (5 doublets and 1 triplet)
 indicating each cluster corresponds to a common ``point source'' in the sky.\cite{cluster}

 However,   within 100 Mpc, we don't find any corresponding
  astronomical objects such as active galactic nuclei, radio galaxy lobes and gamma ray bursts
 which could be a plausible source of the super-GZK or cluster events.\cite{nocandidate}
  In order to circumvent the difficulties in the astrophysical models, several particle physics   
  oriented models have been proposed for the origin of EHECRs.\cite{review} 
  They are the decay of super-heavy relic particles in the galactic halo, the interaction of EHE neutrinos with the cosmic neutrino background in the local cluster.  Violation of Lorentz invariance at the extremely high $\gamma$  factor  is also proposed.

\subsection{Things are not quite that easy}
The HiRes group, on the other hand, recently reported an energy spectrum which is consistent with the existence of the GZK cut-off.\cite{hires}
 After adjusting the energy scale by an amount of 
  error of each experiment($\sim$18\% for AGASA and $\sim$25\% for HiRes),
  the spectra of both experiments agree well below $10^{20}$ eV, 
  yet leave more than a $ \sim 2\sigma$ level  disagreement on the existence of GZK cutoff. 
  We should note that the AGASA uses surface detectors and  the HiRes uses
  fluorescence detectors; both have  drawback and advantage.
    The best way to resolve this contradiction is to measure the air shower simultaneously with an AGASA type air shower array and a HiRes type air fluorescence telescope.

\section{The TA experiment}
In view of these circumstances, 
the hybrid TA was proposed as the first step of building the full TA.  The first step was
financially approved in late  2002 in Japan and the construction was started in 
the Utah desert in collaboration with the US side.
The present TA focuses  whether the EHECR spectrum continues or ends at the GZK energy and whether the discrepancy between AGASA and HiRes is from the statistics, the systematics or the physics. 

The configuration of the  hybrid TA is shown in Fig. 1.
The site is located at  latitude $\sim$39.3$^\circ$N, longitude $\sim$112.8$^\circ$W, and 
the elevation is 
$\sim$1400m above sea level ($\sim$875 g/cm$^2$ atmospheric depth).

 The ground detector consists of an array of 576 counters each of which contains
 2 layers of a 1.2 cm thick plastic scintillator plate of 3 m$^2$ .  They are
 deployed in a grid of 1.2 km spacing covering the ground area of $\sim 760$ km$^2$. The detector acceptance is approximately 9 times that of AGASA. The detection efficiency is 100 \% 
 for cosmic rays with energy more than $10^{19.5}$ eV within zenith angles  45$^\circ$.
   The fluorescence measurement is made at 3 stations surrounding the ground array. The stations form a triangle with a separation of 30 - 40 km. Twelve reflecting telescopes are installed at each station and covers the sky of 3$^\circ$-34$^\circ$ in elevation and 108$^\circ$ in azimuth looking toward the center of the ground array. The diameter of the telescope mirror is $\sim$3.3 m and the pixel resolution
 is approximately $1^\circ\times 1^\circ$. The stereo acceptance is $\sim$670 km$^2$ sr for $E > 10^{20}$  eV by requiring at least one station is within 45km from the shower center. 
 The fluorescence acceptance is 4 times that of AGASA assuming 10\% 
duty factor for observation. 

\begin{figure}
\centerline{\epsfig{file=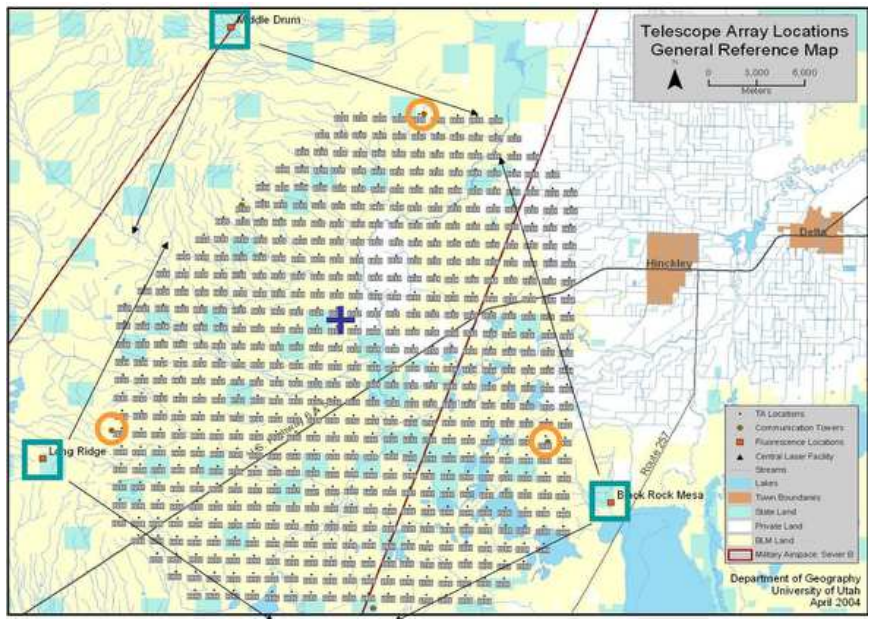, height=8.2cm}}
\caption{Detector Arrangement of Telescope Array. Three square boxes indicate the location of fluorescence telescope stations overlooking a scintillation counter array  indicated by
a number of small boxes.
The circles indicate the locations of communication tower, by which all the controls and data acquisition are relayed to the operation center in the nearby town of Delta, which is
at the rightmost shadow area on the map.}
\end{figure}

\subsection{A bit more detail, current status and prospect}

The surface detectors of  hybrid TA uses thin plastic scintillator. 
Since  the  electro-magnetic component of  the air shower dominates
second dominant  component, i.e, muons more than order of magnitude at least
at  distances not very far from the air shower core, thin detector response 
is governed by the e-m component. This means 
the energy measurement of TA is less affected by the difference of the primary composition and the detail of unknown hadronic interactions at EHE.

That is, the e-m component is governed by the energetic  forward 
particles which are less affected by the nucleus effect, while
muons
at EHE come only from the backward hemisphere of the CMS where nucleus
effect is very large,
 and forward pions and kaons must wait for  the 2nd, 3rd ...  collisions
until their product can copiously decay into muons.

This is the advantage with respect to the use of water Cherenkov counter, which is equally sensitive to the high energy muons and soft gamma rays. We expect the uncertainty of the absolute energy scale can be tuned  better than10 \%
accuracy by carefully analyzing the  simultaneously measured events for 
$E > 10^{19}$ eV. All together, an energy spectrum with twice smaller statistical error than the present AGASA will be obtained by the 3-year measurement of the array and the telescope. 

We use flush ADC and  able to record  the full wave form of
PMT signal.  Therefore, 
the angular resolution of the TA ground array  is expected to be
$\sim 1^\circ$ which is better than AGASA's  $1.6^\circ$.
Moreover,  
the angular resolution expected for the fluorescence and hybrid events are significantly better and they contribute for identifying the cluster events particularly at
 energies  below $10^{19.6}$ eV.  It may be noted that the energy spectrum from the clusters  shows a difference from the general spectrum and can be given a
 plausible reasoning.\cite{Parizot}
 
   In addition, the magnetic deflection by the galactic field is smaller and more regular  in  the  northern  sky,  which  will be  advantageous  to  confirm  the cluster event of AGASA and to
search for corresponding astronomical sources. If the primary EHECRs are charged particles, systematic magnetic deflections with respect to the direction of the galactic field would be observed, which may be used for determining the chemical composition of the primary cosmic rays.  Some of the EHECR source models also predict a large scale anisotropy of arrival directions. 

For identifying the origin of EHECRs, it is essential to identify the particle species of the primary cosmic ray. For example, many of the particle physics oriented models predict an abundant generation of EHE gamma rays and neutrinos rather than protons. On the other hand, the conventional shock wave acceleration will be strongly supported if heavy nuclei are identified as the major composition of EHECRs. For the TA, the particle identification is provided by the shower profile measurement by the fluorescence telescopes. 
The Landau-Pmenranchuk-Migdal 
and geomagnetic effects on the photon primary showers over
 $\sim 10^{19}$ eV give us an interesting clue to  
 identify  EHE photons.\cite{kk}

The EHE neutrinos produce nearly horizontal showers and are easy to identify. The target volume of the first phase TA for EHE neutrino is $\sim 10^{10}$
ton sr and the effective neutrino aperture is $\sim 0.03$ km$^2$sr 
for energies above $10^{20}$ eV. 
The expected neutrino event rate is, however,  $0.04\sim 2$ events
in 10 years, so that we will have to wait for a future deployment of the full TA or the AGASA$\times 100$ large ground arrays.

The mass deployment of detectors
will start in early 2006 and will be completed by March 2007. 
The LHC calibration of the interaction models and liniac calibration of
fluorescence detection will help to draw   
the decisive conclusion on the GZK region spectrum by 2010.

Details of the TA detector design and performance as well as the proposal for low energy extension (TALE) are in the proceedings of the International Cosmic Ray Conference,
2005 (India).

\section*{Acknowledgments}
The Telescope Array is being constructed by the support of Grant-in-Aid for Scientific Research (Kakenhi) on the Priority Area  ``The Highest Energy Cosmic Rays''
 by the Ministry of Education, Culture, Sports, Science and Technology of Japan.

\end{document}